\documentclass[prl,aps,twocolumn,floatfix,showpacs]{revtex4}
\usepackage{graphics}
\begin{document}
\title{Asymmetric two-component Fermi gas with unequal masses}
\author{M. Iskin and C. A. R. S{\'a} de Melo}
\affiliation{School of Physics, Georgia Institute of Technology, Atlanta, Georgia 30332, USA}
\date{\today}

\begin{abstract}
We analyze the zero temperature phase diagram for an asymmetric 
two-component Fermi gas as a function of mass anisotropy and population imbalance.
We identify regions corresponding to normal, or uniform/non-uniform 
superfluid phases, and discuss topological quantum phase transitions 
in the Bardeen-Cooper-Schrieffer (BCS), unitarity 
and Bose-Einstein condensation (BEC) limits.
Lastly, we derive the zero temperature low frequency and long wavelength
collective excitation spectrum, and recover the Bogoliubov 
relation for weakly interacting dilute bosons in the BEC limit. 
\end{abstract}
\maketitle

The evolution from Bardeen-Cooper-Schrieffer (BCS) to Bose-Einstein condensation (BEC)
is a very important topic of current research for the condensed matter, nuclear, atomic and molecular
physics communities. 
Recent advances in atomic physics have allowed for the study of superfluid properties 
in symmetric two-component fermion superfluids (equal mass and equal population) as a function 
of scattering length, where the theoretically predicted crossover from BCS to BEC was observed~\cite{leggett,nsr}.

Since the population of each component as well as the interaction strength between two components 
are experimentally tunable, these knobs enabled the study of the BCS to BEC evolution in 
asymmetric two-component fermion superfluids (equal mass but unequal population)~\cite{mit,rice}.  
In contrast with the crossover physics found in the symmetric case,
these experiments have demonstrated the existence of phase transitions between  
normal and superfluid phases, as well as phase separation between superfluid (paired)
and normal (excess) fermions as a function of population imbalance.

Arguably one of the next frontiers of exploration in cold Fermi gases
is the study of asymmetric two-component fermion superfluidity 
(unequal masses and equal or unequal population)
in two-species fermion mixtures from the BCS to the BEC limit.
Earlier works on two-species fermion mixtures were limited to the
BCS regime~\cite{liu,yip-stability,bedaque,lianyi}. However, very recently, 
the evolution from BCS to BEC was preliminarily addressed in homogenous systems 
as a function of population imbalance and scattering length 
for $^6$Li and $^{40}$K mixture~\cite{iskin-mixture}. 
In addition, the superfluid phase diagram of trapped systems 
at unitarity was also analyzed as a function of population imbalance 
and mass anisotropy~\cite{pao-mixture}.

In this manuscript, we study the BCS to BEC evolution of 
asymmetric two-component fermion superfluids
as a function of population imbalance and mass anisotropy.
Our results for a homogeneous system are as follows.
We analyze the zero temperature phase diagram for an asymmetric 
two-component Fermi gas as a function of mass anisotropy and population imbalance.
We identify regions corresponding to normal, and uniform or non-uniform 
superfluid phases, and discuss topological quantum phase transitions 
in the BCS, unitarity and BEC limits.
Lastly, we derive the zero temperature low frequency and long wavelength
collective excitation spectrum, and recover the Bogoliubov 
relation for weakly interacting dilute bosons in the BEC limit.

To describe a dilute asymmetric two-component Fermi gas in three dimensions, 
we start from the pseudo-spin singlet Hamiltonian ($\hbar = 1$)
\begin{equation}
\label{eqn:hamiltonian}
H = \sum_{\mathbf{k},\sigma} \xi_{\mathbf{k},\sigma} a_{\mathbf{k},\sigma}^\dagger a_{\mathbf{k},\sigma} + 
\sum_{\mathbf{k},\mathbf{k'},\mathbf{q}} V (\mathbf{k},\mathbf{k'})
b_{\mathbf{k},\mathbf{q}}^\dagger b_{\mathbf{k'},\mathbf{q}}, 
\end{equation}
where the pseudo-spin $\sigma$ labels the hyperfine states
represented by the creation operator $ a_{\mathbf{k},\sigma}^\dagger$, and
$b_{\mathbf{k},\mathbf{q}}^\dagger = a_{\mathbf{k}+\mathbf{q}/2,\uparrow}^\dagger 
a_{-\mathbf{k}+\mathbf{q}/2,\downarrow}^\dagger$.
Here, $\xi_{\mathbf{k},\sigma}= \epsilon_{\mathbf{k},\sigma} - \mu_\sigma$,
where $\epsilon_{\mathbf{k},\sigma} = k^2/(2m_\sigma)$ is the energy
and $\mu_\sigma$ is the chemical potential of the fermions.
Notice that, we allow for the fermions to have different masses $m_{\sigma}$ and
different populations controlled by independent chemical potentials $\mu_{\sigma}$.
The attractive fermion-fermion interaction $V (\mathbf{k},\mathbf{k'})$ 
can be written in a separable form as
$
V (\mathbf{k},\mathbf{k'}) =  - g \Gamma^*_\mathbf{k} \Gamma_\mathbf{k'} 
$
where $g > 0$, and $\Gamma_\mathbf{k} = 1$ for the s-wave contact interaction
considered in this manuscript.

The gaussian effective action for $H$ is
$
S_{\rm gauss} = S_0 + (\beta/2) \sum_{q} \bar{\Lambda}^\dagger(q) \mathbf{F}^{-1}(q) \bar{\Lambda}(q),
$
where $q=(\mathbf{q},v_\ell)$ with bosonic Matsubara frequency $v_\ell=2\ell\pi/\beta$.
Here, $\beta = 1/T$, $\bar{\Lambda}^\dagger(q)$ is the order parameter fluctuation field, 
and the matrix $\mathbf{F}^{-1}(q)$ is the inverse fluctuation propagator. 
The saddle point action is
\begin{eqnarray}
S_0 = \beta \frac{|\Delta_0|^2}{g} &+& \sum_\mathbf{k}\big\lbrace
\beta(\xi_{\mathbf{k},+} - E_{\mathbf{k},+}) \nonumber \\
&+& 
\ln [n_F(-E_{\mathbf{k},\uparrow})] + \ln [n_F(-E_{\mathbf{k},\downarrow})] 
\big\rbrace, 
\end{eqnarray}
where
$
E_{\mathbf{k},\sigma} = (\xi_{\mathbf{k},+}^2 + |\Delta_\mathbf{k}|^2)^{1/2} + s_\sigma \xi_{\mathbf{k},-}
$
is the quasiparticle energy when $s_\uparrow = 1$ or
the negative of the quasihole energy when $s_\downarrow = -1$, and
$
E_{\mathbf{k},\pm} = (E_{\mathbf{k},\uparrow} \pm E_{\mathbf{k},\downarrow})/2.
$
Here,
$
\Delta_\mathbf{k} = \Delta_0\Gamma_\mathbf{k}
$
is the order parameter,
$
n_F(E_{\mathbf{k},\sigma})
$
is the Fermi distribution and
$
\xi_{\mathbf{k},\pm} = (\xi_{\mathbf{k},\uparrow} \pm \xi_{\mathbf{k},\downarrow})/2
= k^2/(2m_\pm) - \mu_\pm,
$
where 
$
m_\pm = 2m_\uparrow m_\downarrow/(m_\downarrow \pm m_\uparrow)
$
and 
$
\mu_\pm = (\mu_\uparrow \pm \mu_\downarrow)/2.
$
Notice that $m_+$ is twice the reduced mass of the $\uparrow$ 
and $\downarrow$ fermions, and that the equal mass case corresponds 
to $|m_-| \to \infty$. 
The fluctuation term in the action leads to a correction
to the thermodynamic potential, which can be written as 
$\Omega_{{\rm gauss}} = \Omega_0 + \Omega_{{\rm fluct}}$ with 
$\Omega_0 = S_0/\beta$ and
$\Omega_{{\rm fluct}} = (1/\beta)\sum_{q}\ln\det[\mathbf{F}^{-1}(q)/\beta]$.

The saddle point condition $\delta S_0 /\delta \Delta_0^* = 0$ leads 
to an equation for the order parameter
\begin{equation}
\frac{1}{g} = \sum_{\mathbf{k}} \frac{|\Gamma_\mathbf{k}|^2} {2E_{\mathbf{k},+}} {\cal X}_{\mathbf{k},+},
\end{equation}
where
$
{\cal X}_{\mathbf{k},\pm} = ( {\cal X}_{\mathbf{k},\uparrow} \pm {\cal X}_{\mathbf{k},\downarrow} ) / 2
$
with
$
{\cal X}_{\mathbf{k},\sigma} = \tanh(\beta E_{\mathbf{k},\sigma}/2).
$
As usual, we eliminate $g$ in favor of the scattering length $a_F$ via the relation
$
1/g = - m_+ V /(4\pi a_F) + \sum_{\mathbf{k}} |\Gamma_\mathbf{k}|^2 / (2\epsilon_{\mathbf{k},+}),
$
where
$
\epsilon_{\mathbf{k},\pm} = (\epsilon_{\mathbf{k},\uparrow} \pm \epsilon_{\mathbf{k},\downarrow})/2.
$
The order parameter equation has to be solved self-consistently with number equations
$N_\sigma = -\partial \Omega/\partial {\mu_\sigma}$ which have two contributions
$N_\sigma = N_{0,\sigma} + N_{{\rm fluct},\sigma}$.
$N_{0,\sigma}$ is the saddle point number equation given by
\begin{equation}
N_{0,\sigma} = - \frac{\partial \Omega_0} {\partial {\mu_\sigma}} 
= \sum_{\mathbf{k}} 
\left( \frac{1 - s_\sigma {\cal X}_{\mathbf{k},-}} {2}
- \frac{\xi_{\mathbf{k},+}}{2E_{\mathbf{k},+}} {\cal X}_{\mathbf{k},+}
\right)
\label{eqn:numbereqn}
\end{equation}
and $N_{{\rm fluct},\sigma} = -\partial \Omega_{{\rm fluct}}/\partial {\mu_\sigma}$ is the 
fluctuation contribution to $N$ given by
$
N_{{\rm fluct},\sigma} = - (1/\beta) \sum_{q} \lbrace 
\partial [\det \mathbf{F}^{-1}(q)] / \partial {\mu_\sigma} 
\rbrace / \det \mathbf{F}^{-1}(q).
$

In order to analyze the phase diagram at $T = 0$, we solve the saddle point 
self-consistency (order parameter and number) equations
as a function of population imbalance $P = N_-/N_+$ and 
mass anisotropy $m_r = m_\uparrow / m_\downarrow$.
Here, $N_\pm = (N_\uparrow \pm N_\downarrow)/2$ corresponding to the following 
momenta $k_{F,\pm}^3 = (k_{F,\uparrow}^3 \pm k_{F,\downarrow}^3)/2$.
In addition, we check the stability of saddle point solutions for the
uniform superfluid phase using two criteria.
The first criterion requires that the curvature
$\partial^2 \Omega_0 / \partial \Delta_0^2$ of the saddle point thermodynamic
potential $\Omega_0$ with respect to the saddle point parameter $\Delta_0$ to be 
positive, where 
\begin{equation}
\frac{\partial^2 \Omega_0} {\partial \Delta_0^2} = 
\sum_{\mathbf{k}} |\Delta_{\mathbf{k}}|^2 
\left( 
\frac{{\cal X}_{\mathbf{k}, +}}{E_{\mathbf{k},+}^3} -
\beta \frac{{\cal Y}_{\mathbf{k}, +}}{2E_{\mathbf{k},+}^2}  
\right).
\label{eqn:curvature}
\end{equation}
Here,
$
{\cal Y}_{\mathbf{k},\pm} = ({\cal Y}_{\mathbf{k},\uparrow} \pm  {\cal Y}_{\mathbf{k},\downarrow})/2
$
with ${\cal Y}_{\mathbf{k},\sigma} = {\rm sech}^2 (\beta E_{ \mathbf {k}, \sigma}/2)$. 
When $\partial^2 \Omega_0 /\partial \Delta_0^2$ is
negative, the uniform saddle point solution does not correspond to a minimum of $\Omega_0$, 
and a non-uniform superfluid phase is favored.
This criterion is related to the positive definitenes criterion
on the compressibility matrix $\mathbf{\kappa}$ with elements 
$\kappa_{\sigma,\sigma'} = - \partial^2 \Omega_0 / (\partial \mu_\sigma \partial \mu_{\sigma'})$.
The second criterion requires that the superfluid density
\begin{equation}
\rho_0 (T) = m_\uparrow N_\uparrow + m_\downarrow N_\downarrow
- \frac{\beta}{6} \sum_{\mathbf{k}} k^2 {\cal Y}_{\mathbf{k},+}
\end{equation}
to be positive.
When $\rho_0 (T)$ is negative, a spontaneously generated gradient
of the phase of the order parameter appears, leading to a non-uniform superfluid phase.

\begin{figure} [htb]
\centerline{\scalebox{0.5}{\includegraphics{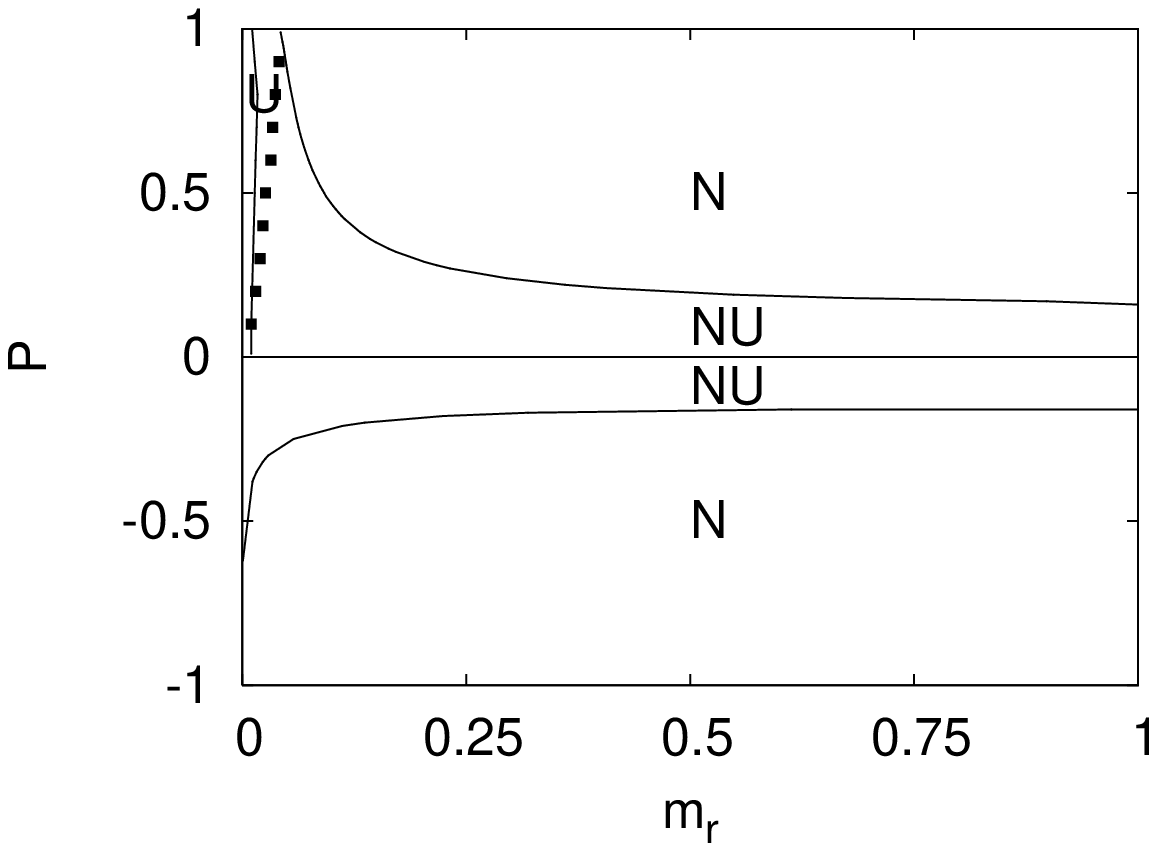} }}
\centerline{\scalebox{0.5}{\includegraphics{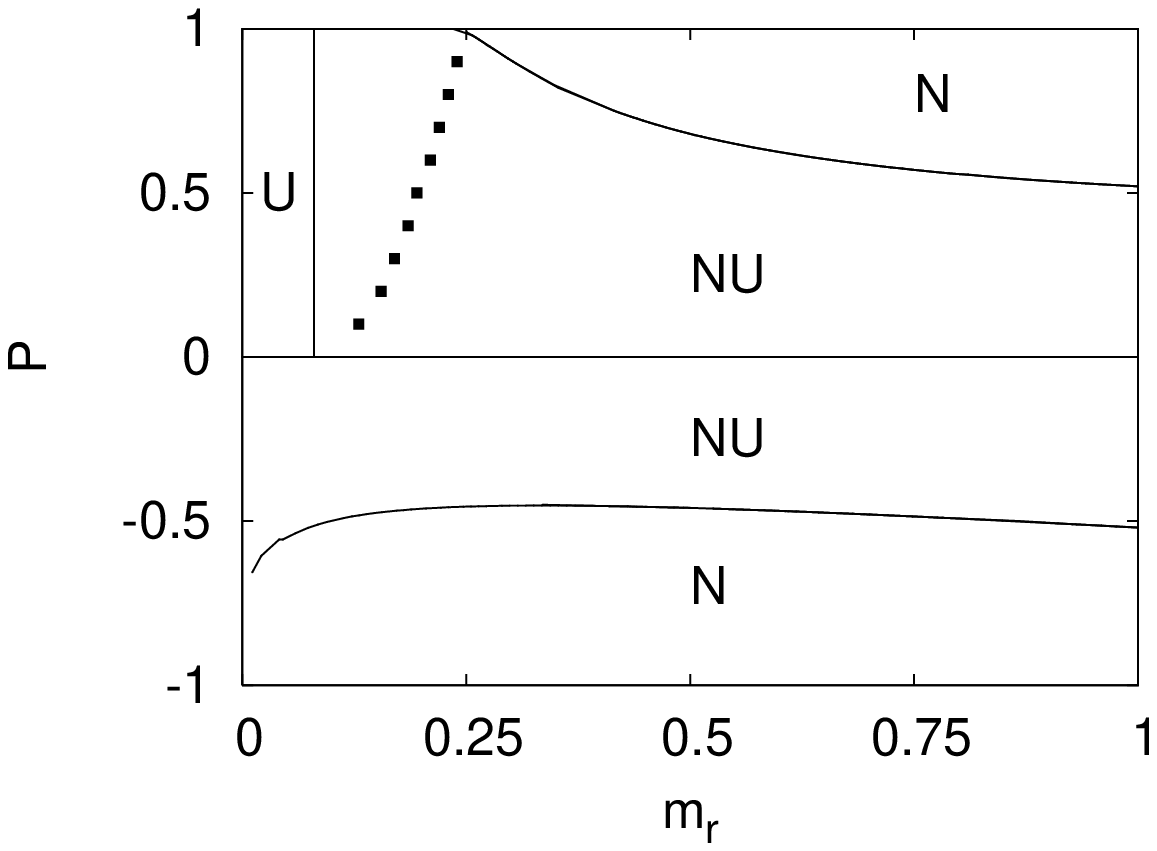} }}
\caption{\label{fig:phase.bcs} 
Phase diagram of $P = (N_\uparrow - N_\downarrow)/(N_\uparrow + N_\downarrow)$ 
versus $m_r = m_\uparrow / m_\downarrow$ on the BCS side 
when a) $1/(k_{F,+} a_F) = -1$ and b) $1/(k_{F,+} a_F) = -0.25$.
}
\end{figure}

\begin{figure} [htb]
\centerline{\scalebox{0.5}{\includegraphics{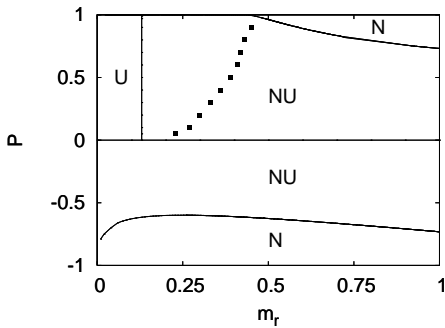} }}
\caption{\label{fig:phase.unitarity} 
Phase diagram of $P = (N_\uparrow - N_\downarrow)/(N_\uparrow + N_\downarrow)$ 
versus $m_r = m_\uparrow / m_\downarrow$ at the unitarity
when $1/(k_{F,+} a_F) = 0$.
}
\end{figure}

\begin{figure} [htb]
\centerline{\scalebox{0.5}{\includegraphics{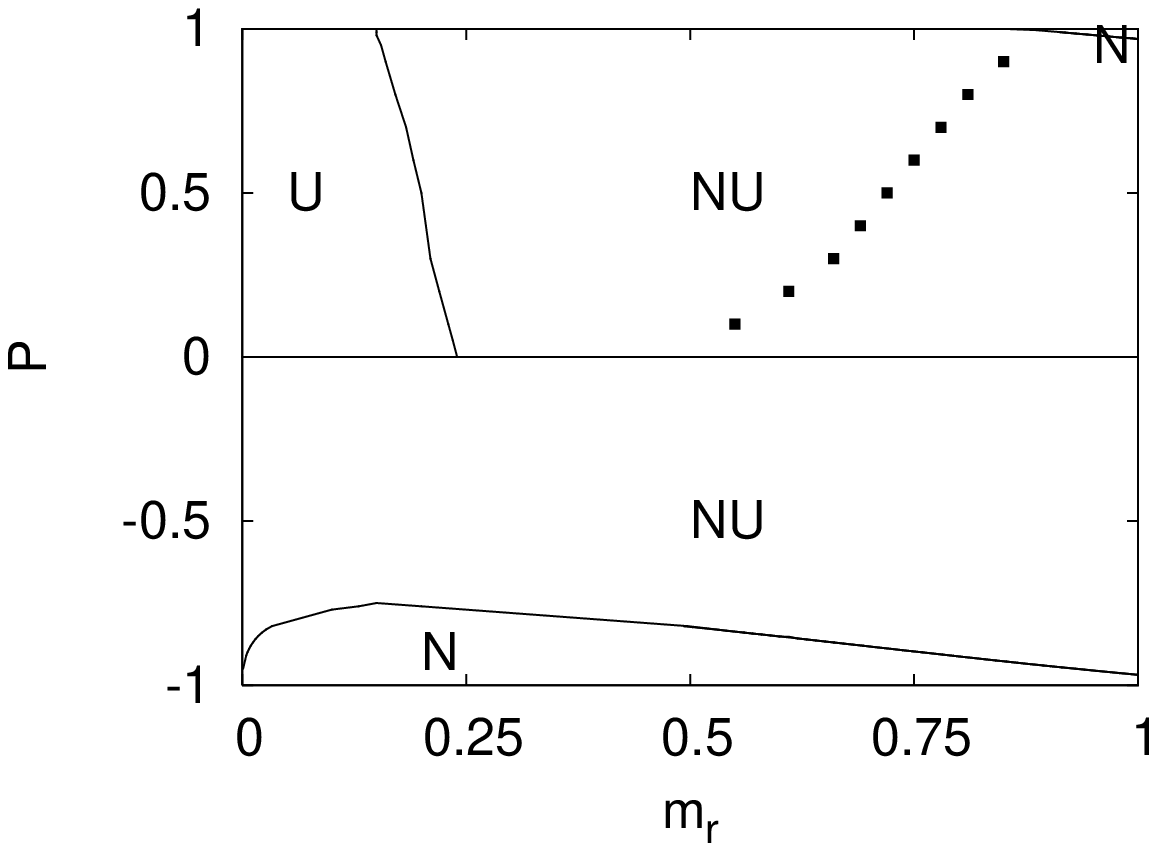} }}
\centerline{\scalebox{0.5}{\includegraphics{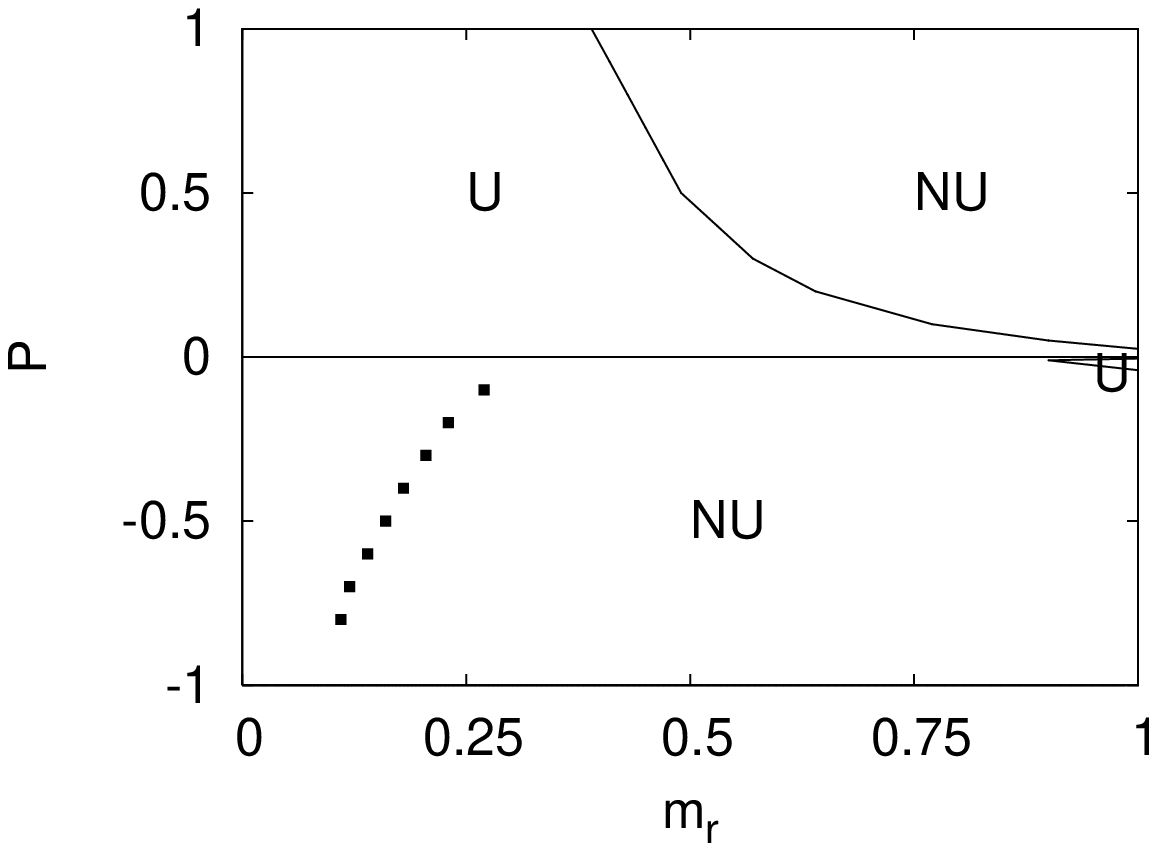} }}
\caption{\label{fig:phase.bec} 
Phase diagram of $P = (N_\uparrow - N_\downarrow)/(N_\uparrow + N_\downarrow)$ 
versus $m_r = m_\uparrow / m_\downarrow$ on the BEC side 
when a) $1/(k_{F,+} a_F) = 0.25$ and b) $1/(k_{F,+} a_F) = 1$.
}
\end{figure}

Based on these two criteria, we construct the $P$ versus $m_r$ phase diagram
for five sets of interaction strengths:
$1/(k_{F,+} a_F) = -1$ and $-0.25$ on the BCS side shown in Fig.~\ref{fig:phase.bcs}; 
$1/(k_{F,+} a_F) = 0$ at unitarity shown in Fig.~\ref{fig:phase.unitarity}; and 
$1/(k_{F,+} a_F) = 0.25$ and $1$ on the BEC side shown in Fig.~\ref{fig:phase.bec}.
In these diagrams, the $\uparrow$ ($\downarrow$) label 
always corresponds to lighter (heavier) mass such that lighter (heavier) fermions
are in excess when $P > 0$ $(P < 0)$. Notice that this choice spans all possible population 
imbalances and mass ratios. In Figs.~\ref{fig:phase.bcs}-\ref{fig:phase.bec}, 
we indicate the regions of normal (N), and non-uniform (NU) or uniform (U)
superfluid phases. 
The normal phase is characterized by a vanishing order parameter ($\Delta_0 = 0$),
while the uniform superfluid phase is characterized by $\rho_0 (0) > 0$ and  
$\partial^2 \Omega_0 /\partial \Delta_0^2 > 0$.
The non-uniform superfluid phase is characterized by
$\rho_0 (0) < 0$ and/or $\partial^2 \Omega_0 /\partial \Delta_0^2 < 0$, and
it should be of the LOFF-type having one wavevector modulation only near the BCS limit, 
although closer to unitarity, we expect the non-uniform phase to be substantially 
different from the LOFF phases having spatial modulation that would encompass several 
wavevectors. However, from numerical calculations, the first criterion seems to be dominant for all parameter 
space and the non-uniform superfluid phase is characterized by
$\partial^2 \Omega_0 /\partial \Delta_0^2 < 0$, which indicates possibly phase separation.

We also identify a topological quantum phase transition, which is shown as dotted lines
in Figs~\ref{fig:phase.bcs}-\ref{fig:phase.bec}.
These phases are characterized by the number of zeros 
of $E_{\mathbf{k},\uparrow}$ and $E_{\mathbf{k},\downarrow}$
(zero energy surfaces in momentum space) such that
I) $E_{\mathbf{k},\sigma}$ has no zeros and $E_{\mathbf{k},-\sigma}$ has only one, and
II) $E_{\mathbf{k},\sigma}$ has no zeros and $E_{\mathbf{k},-\sigma}$ has two zeros.
Phase I (II) always appears to the left (right) of the dotted lines.
Notice that, the $P  = 0$ limit corresponds to case III, 
where $E_{\mathbf{k},\sigma}$ has no zeros and is always positive.
The transition from case II to case I occurring at the dotted lines
is quantum in nature, and signatures of it could still be observed
at finite temperatures through the measurement of momentum 
distributions~\cite{iskin-mixture}. However, phase II seems to lie always 
in the NU region and may not be accessible experimentally.

As shown in Fig.~\ref{fig:phase.bcs}a, we find a small region of uniform
superfluidity on the BCS side for $1/(k_{F,+} a_F) = -1$ only when the mass anisotropy is small and the
lighter fermion are in excess ($P > 0$). 
Thus, mixtures consisting of $^6$Li and $^{40}$K ($m_r \approx 0.15$) 
or $^6$Li and $^{87}$Sr ($m_r \approx 0.07$) are good candidates for future experiments.
In the rest of the phase diagram, 
we find a quantum phase transition from the non-uniform 
superfluid to the normal phase beyond a critical population imbalance 
for both positive and negative $P$.
The phase space of uniform superfluidity expands while that of the 
normal phase shrinks with increasing interaction strength as shown 
in Fig.~\ref{fig:phase.bcs}b. 

This general trend continues into the unitarity limit [$1/(k_{F,+} a_F) = 0$] 
as shown in Fig.~\ref{fig:phase.unitarity}. Since this limit is theoretically 
important as well as experimentally accessible, 
we suggest that Fermi mixtures corresponding to $0 < m_r < 0.45$ have 
phase diagrams which are qualitatively different from those of $ 0.45 < m_r < 1$
as a function of $P$. Thus, mixtures consisting
of $^6$Li and $^2$H ($m_r \approx 0.33$), $^6$Li and $^{25}$Mg ($m_r \approx 0.24$)
or $^{40}$K and $^{87}$Sr ($m_r \approx 0.64$) are also good candidates for future experiments.
Notice that, our results for the case of equal masses ($m_r = 1$) are in close 
agreement with recent MIT experiments~\cite{mit} in a trap.
At unitarity, our non-uniform superfluid to normal state 
boundary occurs at $P \approx \pm 0.73$, and the MIT group obtains 
$P \approx \pm 0.70(4)$ for their superfluid to normal boundary.

Additional increase of interaction strength beyond unitarity on the BEC side 
leads to further expansion (shrinkage) of the uniform superfluid (normal) region
as shown in Fig.~\ref{fig:phase.bec}.
When heavier fermions are in excess ($P < 0$), a uniform superfluid phase 
is not possible for any mass anisotropy until a critical interaction 
strength is reached. The critical interaction strength corresponds to
$1/(k_{F,+} a_F) \approx 0.8$ for $m_r = 1$.
However, in the extreme BEC limit [$1/(k_{F,+} a_F) >> 1$], 
only the uniform superfluid phase exists even for $P < 0$ (not shown).

Next, we analyze gaussian fluctuations from 
which we extract the low frequency and long wavelength collective excitations at zero temperature.
The collective excitation spectrum is determined by the poles of the 
propagator matrix $\mathbf{F} (q)$ determined
by the condition $\det \mathbf{F}^{-1} (q) = 0$, when the usual analytic continuation
$iv_\ell \rightarrow w + i0^+$ is performed.
First, we express the fluctuation field as $\Lambda(q) = [ \lambda(q) +i \theta(q) ] / \sqrt{2}$,
where $\lambda(q)$ and $\theta(q)$ are amplitude and phase fields, respectively.
Then, we consider only the $P = 0$ or 
$k_{F,+} = k_{F,\uparrow} = k_{F,\downarrow}$ limit,
and expand the matrix elements of $\mathbf{F}^{-1}(q)$ 
to second order in $|\mathbf{q}|$ and $w$ to get
\begin{equation}
\mathbf{F}^{-1}(\mathbf{q},w) = \left( \begin{array}{cc} A + C|\mathbf{q}|^2 - Dw^2 & iBw 
\\ -iBw & Q|\mathbf{q}|^2 - Rw^2 \end{array}\right).
\label{eqn:coll}
\end{equation}
Thus, there are two branches for the collective excitations, but we focus only on
the lowest energy one correponding to the Goldstone mode with dispersion
$
w(\mathbf{q}) = v |\mathbf{q}|,
$
where $v = \sqrt{A Q / (AR + B^2)}$ is the speed of sound.
Notice that extra care is required when $P \ne 0$ since Landau damping causes collective excitations 
to decay into the two-quasiparticle continuum even for the s-wave case.

The BCS limit is characterized by the criteria $\mu_+ > 0$ and 
$\mu_+ \approx \epsilon_{F,+} \gg |\Delta_0|$.
The expansion of the matrix elements to order $|{\mathbf q}|^2$ and $w^2$ is
performed under the condition $[w,|\mathbf{q}|^2/(2m_+)] \ll |\Delta_0|$.
The coefficient that couples phase and amplitude fields
vanish ($B = 0$) in this limit. Thus, there
is no mixing between the phase and amplitude modes.
The zeroth order coefficient is
$
A = {\cal D} ,
$
and the second order coefficients are
$
C = Q/3 = {\cal D} v_{F,\uparrow} v_{F,\downarrow}/ (36|\Delta_0|^2),
$
and
$
D = R/3 = {\cal D} / (12|\Delta_0|^2).
$
Here, $v_{F,\sigma} = k_{F,\sigma}/m_\sigma $ is the Fermi velocity and 
${\cal D} = m_+ V k_{F,+}/(2\pi^2)$ is the 
density of states per spin at the Fermi energy.
Thus, we obtain
$
v = \sqrt{v_{F,\uparrow} v_{F,\downarrow} / 3} = \sqrt{v_\uparrow v_\downarrow},
$
with $v_\sigma = v_{F,\sigma}/\sqrt{3}$, 
which reduces to the Anderson-Bogoliubov relation when the masses are equal.

\begin{figure} [htb]
\centerline{\scalebox{0.6}{\includegraphics{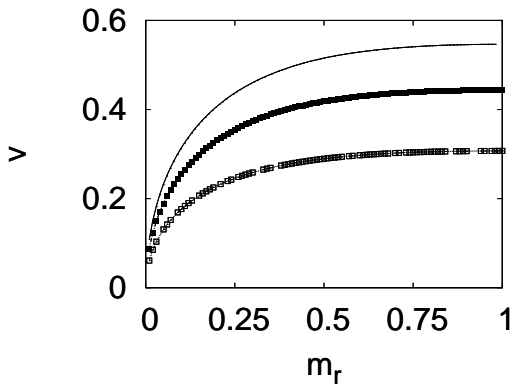} }}
\caption{\label{fig:collective}
Sound velocity $v$ (in units of $v_{F,+} = k_{F,+}/m_+$) versus 
$m_r = m_\uparrow / m_\downarrow$ for
$1/(k_{F,+} a_F) = -1$ (solid line),
$1/(k_{F,+} a_F) = 0$ (solid squares) and 
$1/(k_{F,+} a_F) = 1$ (hollow squares).
Here, populations are equal ($P = 0$).
}
\end{figure}

On the other hand, the BEC limit is characterized by the criteria $\mu_+ < 0$ and 
$\xi_{\mathbf{k},+} \gg |\Delta_0|$.
The expansion of the matrix elements to order $|{\mathbf q}|^2$ and $w^2$ is
performed under the condition $[w, |\mathbf{q}|^2/(2m_+)] \ll |\mu_+|$.
The coefficient $B \ne 0$ indicates that the amplitude and phase fields are mixed.
The zeroth order coefficient is
$
A = \kappa |\Delta_0|^2 / (2|\mu_+|),
$
the first order coefficient is
$
B = \kappa,
$
and the second order coefficients are
$
C = Q = \kappa/[2(m_\uparrow + m_\downarrow)]
$
and
$
D = R = \kappa /(8|\mu_+|),
$
where
$
\kappa = {\cal D} / (32\sqrt{|\mu_+|\epsilon_{F,+}}).
$
Thus, we obtain
$
v = |\Delta_0| / \sqrt{4(m_\uparrow + m_\downarrow) |\mu_+|}
= \sqrt{v_\uparrow v_\downarrow}
$
with $v_\sigma = \sqrt{2\pi n_\sigma a_F / m_\sigma^2}$.
Notice that the sound velocity is very small and its smallness is controlled
by the scattering length $a_F$.
Furthermore, in the theory of weakly interacting dilute Bose gas, 
the sound velocity is given by 
$
v_B = \sqrt{4\pi a_{BB} n_B / m_B^2}.
$
Making the identification that the density of 
pairs is $n_B = n_+$, the mass of the bound pairs is 
$m_B = m_\uparrow + m_\downarrow$ and that the Bose scattering length is 
$
a_{BB} = (m_B/m_+) a_F = [1 + m_\uparrow/(2m_\downarrow) + m_\downarrow/(2m_\uparrow)] a_F, 
$
$v_B$ reduces to the well known Bogoliubov relation when the masses are equal.
Therefore, the strongly interacting Fermi gas with two species can be described as 
a weakly interacting Bose gas at zero temperature as well as at finite temperatures~\cite{iskin-mixture}.
Notice that $a_{BB}$ reduces to $a_{BB} = 2a_F$ for equal masses~\cite{jan} in the
Born approximation, but a better estimate for $a_{BB}$ can be found in the literature~\cite{gora}.

In Fig.~\ref{fig:collective}, we show the sound velocity as a function of the
mass ratio $m_r$ for three values of the scattering parameter $1/(k_{F,+} a_F) = -1, 0$ and $1$ corresponding
to the BCS side [$1/(k_{F,+} a_F) = -1$], unitarity [$1/(k_{F,+} a_F) = 0$], and to the BEC side 
[$1/(k_{F,+} a_F) = 1$]. Notice that the speed of sound could be measured for a given $m_r$ 
using similar techniques as in the single species case $m_r = 1$~\cite{thomas,grimm}.

In summary, we analyzed the zero temperature phase diagram for an asymmetric
two-component Fermi gas as a function of mass anisotropy and population imbalance.
We identified regions corresponding to normal, and uniform or non-uniform
superfluid phases, and discussed topological quantum phase transitions
in the BCS, unitarity and BEC limits.
Lastly, we derived the zero temperature low frequency and long wavelength
collective excitation spectrum, and recovered the Bogoliubov 
relation for weakly interacting dilute bosons in the BEC limit.
We thank NSF (DMR-0304380) for support.

Note added: We would like to acknowledge valuable e-mail 
correspondence with S.-K. Yip from June to August of 2006
in connection with differences and similarities between the 
stability criteria for the phase diagrams of the original versions of references~\cite{pao-mixture}
and~\cite{iskin-mixture} as well as of the present manuscript. In addition, recently, a 
few manuscripts have appeared addressing phase diagrams and their
stability criteria for unequal population Fermi systems~\cite{lianyi2, gubankova, sheehy, levin, pao-stability}. 
These manuscripts have 
lead to a debate whether the compressibility criterion and the 
criterion based on the curvature of the thermodynamic potential with respect to the
order parameter are equivalent or not. Our comments in connection to the current 
debate will be posted at a later date.

\end{document}